\documentclass[11pt]{article}

\newtheorem{example}{Example}

\begin{document}

\title{Relativistic Orbits and the Zeros of $\wp(\Theta)$}
\author{K. Huber \\ 
           Huber Consult \\
           Sesenheimer Str. 21 \\
          10627 Berlin \\ 
          Germany         \\
 email: email@klaus-huber.net}
\date{ }
\maketitle
\begin{abstract}
A simple expression for the zeros of Weierstrass' function is given
which follows from a formula for relativistic orbits.

\vspace{1mm}

\end{abstract}

\section{Introduction}

For elliptic functions, Weierstrass' function $\wp(\Theta)$ is a very important tool.
It appears in countless applications in mathematics, physics, and engineering.
There is some interest to determine its zeros. This problem is considered as being difficult.
A formula can be found in~\cite{Eichler:Zagier}, see also~\cite{McKean:Moll}. However, to 
quote~(\cite{McKean:Moll}, p.105): "There seems to be no simple way to express the roots".

In this contribution a simple expression is given for the roots of $\wp(\Theta)=0$.
The formula follows from a formula for relativistic orbits (see~\cite{Huber2010Zarm}
and~\cite{Huber2011mm}), which in turn originated from research in the field of 
Cauer filters~\cite{Huber2011}. We assume that the reader is familiar with elliptic
function theory. We use the notation employed by Lawden~\cite{Lawden}.

\section{Relativistic Orbits}
\label{sec:RelOrb}

In Lawden's book (\cite{Lawden}, pp.126-129), it has been shown, that the 
generalisation of the Kepler ellipse to the relativistic orbit leads to the polar equation
\[ 
r = \frac{1}{A + B \cdot sn^2(C \Theta,k)} \mbox{ , }
\]
where $r$ gives the distance of two bodies and $\Theta$ the angle in the plane of motion 
with suitable reference. 
As Lawden was mainly interested in deriving the perihel shift of the planet Mercury, 
he only determined approximations of the constants $A$, $B$, and $C$. 

\noindent In~\cite{Huber2010Zarm} and~\cite{Huber2011} these constants have been determined
exactly by explicitly solving the orbit differential equation for $u=1/r$ given by
\[ 
u'' + u = \alpha + 3 \beta \cdot u^2 \mbox{.}
\]
The dash denotes derivation with respect to the angle $\Theta$. The constants $\alpha$ and $\beta$ depend on the motion 
to be considered (e.g. planetary motion, photonic motion, motion of elementary particles, etc.).
This led to
\begin{eqnarray}
\nonumber
A & = & \frac{1}{6 \beta} \cdot \left( 
                                               1 - \sqrt{ \frac{1-12 \alpha \beta}{k^4-k^2+1}} (k^2+1)
                                            \right)
            \\
\nonumber
B & = & \frac{k^2}{2 \beta}\cdot \sqrt{ \frac{1-12 \alpha \beta}{k^4-k^2+1}} \\
\label{eq:C}
C & = & \frac{1}{2}\cdot \sqrt[4]{ \frac{1-12 \alpha \beta}{k^4-k^2+1}} \mbox{ . }
\end{eqnarray}
Note that the constant $C$ is invariant to the transformation $k^2 \rightarrow k'^2=1-k^2$.

\noindent The shape for the relativistic motion (and for the magnitude of Cauerfilters) has two
relevant constants, the value of $k$ and the value of the product $\alpha \beta$. 
These two parameters essentially determine the trajectories (up to scaling).

\section{Simple Expression for Zeros of  $\wp$}
\label{sec:SimEx}

The orbit equation can be easily solved for the angle $\Theta =(1/C) \cdot sn^{-1}(\sqrt{\frac{1/r-A}{B}})$.
We now show the relation of the orbit equation to the Weierstrass function $\wp$.
To this end we use the relation of $\wp$ to the $sn^2$-function (see~\cite{Smirnow}, p. 553)
\[
 \wp(\Theta) = \frac{e_1-e_3}{sn^2(\sqrt{e_1-e_3} \Theta)} + e_3 \mbox{ , }
\]
or
\begin{equation}
\label{eq:Smirnow}
 \wp(\Theta) = \frac{C^2}{sn^2(C \Theta)} - \frac{(1+k^2) \cdot C^2}{3} \mbox{ , }
\end{equation}
i.e. $C \stackrel{!}{=} \sqrt{e_1-e_3}$ and $e_3=- \frac{1+k^2}{3}C^2$. The value $C$ is 
given by equation~(\ref{eq:C}). The angle $\Theta$ then can be expressed by the integral
\[ 
 \Theta = \int_{\infty}^{\wp(\Theta)} \frac{dy}{\sqrt{4 y^3 - g_2 \, y - g_3}} \mbox{ , }
\]
where $g_2$ and $g_3$ are the Weierstrass invariants which follow from the differential equation
\[
 \wp'^2  = 4 \wp^3 - g_2 \, \wp -g_3
\]
and can be expressed as sum over all non-zero reciprocal 4-th and 6-th powers respectively
of the lattice points of the adjoint parallelograms. The polynomial $4 y^3 - g_2 \, y - g_3$ factors as
\[
4 y^3 - g_2 \, y - g_3 = 4 \, (y-e_1) (y-e_2) (y-e_3) \mbox{ . }
\]
Thus with $C = \sqrt{e_1-e_3}$, $e_3=- \frac{1+k^2}{3}C^2$, and $e_1+e_2+e_3=0$ we get
\begin{eqnarray*}
e_1 & = & -\frac{k^2 - 2}{3} \, C^2 \\
e_2 & = &  \frac{2 k^2-1}{3} \, C^2 \\
e_3 & = & - \frac{k^2+1}{3}  \, C^2 \mbox{ , }
\end{eqnarray*}
which leads to  
\begin{eqnarray*}
g_3 & = &  4 \, e_1 \, e_2 \, e_3 \hspace{5mm} \Rightarrow \hspace{3mm} g_3 = \frac{4}{27}\, (k^2+1) (k^2-2) (2k^2-1) \, C^6 \\
g_2 & = &  -4 \, (e_1 e_2 + e_1 e_3 + e_2 e_3) \Rightarrow g_2 =  \frac{4}{3} (k^4-k^2+1) \, C^4  \mbox{ . }
\end{eqnarray*}
Using equation~(\ref{eq:C}) we obtain the remarkably simple result
\begin{equation}
 g_2 = \frac{1}{12}- \alpha  \beta  \mbox{ , }
\end{equation}
and 
\[ 
 g_3 = 
           \frac{(k^2+1) (k^2-2) (2k^2-1)}{2^4 \cdot 3^3} \, \left( \frac{1 - 12 \alpha \beta}{k^4-k^2+1} \right)^{3/2}  \mbox{ . }
\] 

To obtain a compact description of all cases we compute the discriminant of the polynomial $4 y^3-g_2 y -g_3$ which
is given by $g_2^3-27 g_3^2 = 16 \, (e_1-e_2)^2 (e_1-e_3)^2 (e_2-e_3)^2$, thus 
$ g_2^3-27 g_3^2= 2^4 (1-k^2)^2 k^4 C^{12}$ which leads to
\[ 
 g_2^3-27 g_3^2 = \frac{(1-k^2)^2 \cdot k^4 \cdot (1-12 \alpha \beta)^3}{2^8 \cdot  (k^4-k^2+1)^3}  \mbox{ . }
\] 
If the discriminant is greater than zero we have three real roots, if it is zero we have three real roots which are
not all distinct and for negative discriminant we have one real and two complex zeros. The absolute invariant 
$g_2^3/(27 g_3^2)$ follows as
\begin{equation}
\label{eq:absinv}
 \frac{g_2^3}{27 \, g_3^2} =  \frac{ (k^4-k^2+1)^3}
                                                       {(k^2+1)^2 \cdot (k^2-2)^2  \cdot (k^2-\frac{1}{2})^2} \mbox{ . }
\end{equation}
The absolute invariant transforms the three cases of the discriminant to the cases that $g_2^3/(27 g_3^2)$
is greater, equal or smaller than $1$. The absolute invariant is an extremely useful tool for characterizing
the ratio of the two periods of the corresponding elliptic function.  It is immediately seen that the absolute
invariant is left unchanged by the transformations $k^2 \rightarrow k'^2= 1 - k^2$ (like $C$) and
$k^2 \rightarrow 1/k^2$.

\noindent The zeros of $\wp$ now follow from equation~(\ref{eq:Smirnow})
\[
 \frac{C^2}{sn^2(C \Theta)} = \frac{(1+k^2) \cdot C^2}{3}
 \hspace{2mm} \Rightarrow  \hspace{2mm}
 \Theta_0 = \pm \frac{1}{C}\cdot sn^{-1}(\sqrt{\frac{3}{1+k^2}},k) \mbox{ . }
\]
hence the zeros of the Weierstrass function are given by the simple expression
\begin{equation}
 \Theta_0 = \pm \sqrt[4]{\frac{4 (k^4-k^2+1)}{3 \, g_2}} \cdot sn^{-1}(\sqrt{\frac{3}{1+k^2}},k) \mbox{ . }
\end{equation}

\noindent We give some examples.

\begin{example}
Computing the length of the famous Lemniscate, one encounters the integral
$ \int \frac{dx}{\sqrt{1-x^4}}$ which can  be transformed to the integral $\int \frac{dy}{\sqrt{4 y^3 -4 y}}$.
Hence $g_2=4$ and $g_3=0$. The values of $k^2$ follow from equation~(\ref{eq:absinv}).
The possible values are $k^2=-1$, $k^2=1/2$, and $k^2=2$. We select the value of $k^2$ in the
interval $[0,1]$, i.e. $k^2=1/2$ and obtain
\[
 \Theta_0= \pm \frac{1}{\sqrt{2}} \, sn^{-1}(\sqrt{2},\frac{1}{\sqrt{2}})  \mbox{ . }
\]
The approximate numeric value in the parallelogramm around zero is given by $\approx \pm 1.3110287771\cdot (1-i)$
Clearly this is the value $\pm \frac{\Gamma(\frac{1}{4})^2}{4 \sqrt{2 \pi}}\cdot (1-i)$.
Note that for $k=i$ we have Gauss' {\em Sinus Lemniscatus} 
$sl(z)=sn(z,i)=\frac{1}{2} \, sd(\sqrt{2}z,\frac{1}{\sqrt{2}})$.
\end{example}

\noindent In table~I data for some values of $k$ used in the following examples are collected. All numeric values
for $\Theta_0$ given in the examples are for the parallelogram around zero.

\begin{example}
Let $g_2=7$, $g_3=3$, and $k=\frac{1}{\sqrt{5}}$ (see table~I). This leads to
\[
  \Theta_0   =           \pm \sqrt{\frac{2}{5}} \cdot sn^{-1}(\sqrt{\frac{5}{2}}, \frac{1}{\sqrt{5}}) 
                 \approx   \pm (1.0496381-i \cdot 0.77781243) \mbox{ . }
\]
\end{example}

\begin{example}
Let $g_2=11$, $g_3=7$, and $k=\frac{\sqrt{2}-1}{\sqrt{2}+1}$. This leads to
\[
  \Theta_0 =\pm \sqrt{2} \, \sqrt[4]{17-12 \sqrt{2}} \cdot sn^{-1}(\sqrt{\frac{3}{2}+\sqrt{2}}, \frac{\sqrt{2}-1}{\sqrt{2}+1}) 
  \approx \pm (0.9270373 + i \cdot 0.6766441) \mbox{ . }
\]
\end{example}

\begin{example}
Let $g_2=15$, $g_3=\sqrt{2} \cdot 7$, and $k=\sqrt{2}-1$. This leads to
\[
  \Theta_0 =\pm \sqrt{\frac{2}{3}} \, \sqrt[4]{3-2 \sqrt{2}} \cdot 
         sn^{-1}( \frac{\sqrt{3}}{2} \sqrt{2+\sqrt{2}}, \sqrt{2}-1) 
   \approx \pm (0.86473386 + i \cdot 0.637892607) \mbox{ . }
\]
\end{example}

\noindent We may also use the property 
\[
 \wp(\lambda \Theta | \frac{g_2}{\lambda^4}, \frac{g_3}{\lambda^6}) = \frac{\wp(\Theta | g_2,g_3)}{\lambda^2} 
\]
to obtain additional solutions from known values of $g_2$ and $g_3$ for any $\lambda \neq 0$.

\begin{example}
Using $\lambda=2^{1/12}$, we get the zeros of the Weierstrass function for the 
case $g_2=15 / 2^{1/3}$ and $g_3=7$ from the previous example as
\[
 \Theta_0 = \pm 2^{1/12} \cdot  \sqrt{\frac{2}{3}} \, \sqrt[4]{3-2 \sqrt{2}} \cdot 
          sn^{-1}( \frac{\sqrt{3}}{2} \sqrt{2+\sqrt{2}}, \sqrt{2}-1)  \mbox{ . }
\]
\end{example}

\noindent In general for given $g_2$ and $g_3$ we can easily compute $k$ from equation~(\ref{eq:absinv}).
Using the substitution $\xi=k^2+1/k^2-1$ we get a polynomial of degree three in $\xi$ which can be solved for
$\xi$ from which $k^2$ and $k$ follows. Namely we get the polynomial
\[
  \xi^3 + a \cdot \xi -a =0 
\]
where~\footnote{Note that $a$ is simply related to 
the j-invariant $1728 \, g_2^3/(g_2^3-27 g_3^2)$.} $a=\frac{27}{4} \cdot \frac{g_2^3}{27 g_3^2-g_2^3}$, which solved for $\xi$ yields
\[
 \xi = \rho     \cdot \sqrt[3]{\frac{a}{2} +  \sqrt{   (\frac{a}{3})^3+ (\frac{a}{2})^2}} +
         \rho^2 \cdot \sqrt[3]{\frac{a}{2} -  \sqrt{   (\frac{a}{3})^3+ (\frac{a}{2})^2}}
\]
with $\rho \in \{ 1, e^{i 2 \pi / 3},e^{i 4 \pi / 3} \}$. Then all values of $k^2$ follow:
\begin{equation}
\label{eq:k2vonxi}
 k^2 = \frac{1}{2} \cdot  (\xi + 1 \pm \sqrt{\xi^2+2 \xi -3}) \mbox{ . }
\end{equation}
If a real solution for $k^2$ in the interval $[0,1]$ exists it is obtained by selecting $\rho=1$
and the minus sign in equation~(\ref{eq:k2vonxi}). 

\vspace{8mm}

{\em Table~I: Data for some invariants} \newline

\vspace{1mm}

\begin{tabular}{|c|c|c|c|c|} \hline
 $k$ & $\frac{(k^4-k^2+1)^3}{(k^2+1)^2 (k^2-2)^2(k^2-\frac{1}{2})^2}$ & $\frac{g_2^3}{27 \cdot g_3^2}$  & $g_2$   & $g_3$  \\ \hline \hline
 $\frac{1}{\sqrt{5}}$                       & $\frac{7^3}{3^5}$             & $\frac{7^3}{27 \cdot 3^2}$                                &  $7$  & $3$   \\ \hline
 $\frac{\sqrt{2}-1}{\sqrt{2}+1}$  & $\frac{11^3}{3^3 7^2}$   & $\frac{11^3}{27 \cdot 7^2}$                              &  $11$ & $7$   \\ \hline
 $\sqrt{2}-1$              & $\frac{5^3}{2 \cdot 7^2}$       & $\frac{(3 \cdot 5)^3}{27\cdot (\sqrt{2} \cdot 7)^2}$  &  $15$ & $\sqrt{2} \cdot 7$  \\  \hline
\end{tabular}

\vspace{8mm}

\section{Conclusion}

The simple and easy to compute expression 
\[
 \Theta_0 = \pm \sqrt[4]{\frac{4 (k^4-k^2+1)}{3 \, g_2}} \cdot sn^{-1}(\sqrt{\frac{3}{1+k^2}},k) 
\]
for the zeros of the Weierstrass elliptic function $\wp$ has been derived, where $k$ is given in
closed form as function of $g_2$ and $g_3$.



\begin{thebibliography}{99}

\bibitem{Eichler:Zagier} M.Eichler, D.Zagier, "On the zeros of the Weierstrass $\wp$-function,
	Mathematische Annalen, Vol. 258, No.4, Dec. 1982, pp.399-407.


\bibitem{Huber2010Zarm} K.Huber, "Relativistische Bahnkurven und Cauer Filter",
              Bremen-Oldenburg relativity seminar, Zarm, Universit\"at Bremen, 16th December, 2010, available at
                                \verb+http://www.researchgate.net/profile/Klaus_Huber/+

\bibitem{Huber2011} K.Huber, "Cauer Filters", BOD, Norderstedt, 2011.

\bibitem{Huber2011mm} K.Huber, "Relativistic Motion", BOD, Norderstedt, 2011.

\bibitem{Lawden} D.F.Lawden, "Elliptic Functions and Applications", Springer-Verlag, 
	Berlin Heidelberg New York, 1989.

\bibitem{McKean:Moll} H.McKean, V.Moll, "Elliptic Curves", Cambridge University Press, 1999.

\bibitem{Smirnow} W.I.Smirnow, "Lehrgang der h\"oheren Mathematik",
	Teil III/2, Deutscher Verlag der Wissenschaften, Berlin 1979, (now Harri Deutsch, Frankfurt).



\end{thebibliography}
\end{document}